\documentclass[10pt,conference]{IEEEtran}
\usepackage{latexsym}
\usepackage{cite}
\usepackage{amssymb}
\usepackage{bm}
\usepackage{amsmath, amssymb}
\usepackage[dvips]{graphics}
\usepackage{graphicx}
\usepackage{psfrag}

\newtheorem{theorem}{Theorem}

\newtheorem{corollary}{Corollary}
\newtheorem{example}{Example}

\DeclareMathOperator\E{E}
\DeclareMathOperator*\maxx{max\vphantom{p}}

\begin{document}
\title{On Directed Information and
Gambling}
\author{
\IEEEauthorblockN{Haim H. Permuter}
\IEEEauthorblockA{
Stanford University \\
Stanford, CA, USA \\
haim1@stanford.edu }\and \IEEEauthorblockN{Young-Han Kim}
\IEEEauthorblockA{
University of California, San Diego\\
La Jolla, CA, USA\\
yhk@ucsd.edu }\and
\IEEEauthorblockN{Tsachy Weissman}
\IEEEauthorblockA{
Stanford University/Technion \\
Stanford, CA, USA/Haifa, Israel \\
tsachy@stanford.edu } }

\maketitle

\begin{abstract}
We study the problem of gambling in horse races with causal side
information and show that Massey's directed information characterizes
the increment in the maximum achievable capital growth rate due to the
availability of side information. This result gives a natural
interpretation of directed information $I(Y^n \to X^n)$ as the amount
of information that $Y^n$ \emph{causally} provides about $X^n$.
Extensions to stock market portfolio strategies and data compression
with causal side information are also discussed.
\end{abstract}


%

\section{Introduction}
Mutual information arises as the canonical answer to a variety of
problems. Most notably, Shannon~\cite{Shannon48} showed that the
capacity $C$, the maximum data rate for reliable communication over a
discrete memoryless channel $p(y|x)$ with input $X$ and output $Y$, is
given by
\begin{equation}
\label{eq:shannon}
C = \max_{p(x)} I(X;Y),
\end{equation}
which leads naturally to the operational interpretation of mutual
information $I(X;Y) = H(X) - H(X|Y)$ as the amount of uncertainty
about $X$ that can be reduced by observation $Y$, or equivalently, the
amount of information $Y$ can provide about $X$. Indeed, mutual
information $I(X;Y)$ plays the central role in Shannon's random coding
argument, because the probability that independently drawn $X^n$ and
$Y^n$ sequences ``look'' as if they were drawn jointly decays
exponentially with exponent $I(X;Y)$. Shannon also proved a dual
result \cite{Shannon60} showing that the minimum compression rate $R$
to satisfy a certain fidelity criterion $D$ between the source $X$ and
its reconstruction $\hat{X}$ is given by $R(D) = \min_{p(\hat{x}|x)}
I(X;\hat{X})$. In another duality result (Lagrange duality this time)
to \eqref{eq:shannon}, Gallager \cite{Gal77} proved the minimax
redundancy theorem, connecting the redundancy of the universal
lossless source code to the capacity of the channel with conditional
distribution described by the set of possible source distributions.

Later on, it was shown that mutual information has also an important
role in problems that are not necessarily related to describing
sources or transferring information through channels. Perhaps the most
lucrative example is the use of mutual information in gambling.

Kelly showed in \cite{Kelly56} that if each horse race outcome can be
represented as an independent and identically distributed (i.i.d.)
copy of a random variable $X$ and the gambler has some side
information $Y$ relevant to the outcome of the race, then under some
conditions on the odds, the mutual information $I(X;Y)$ captures the
difference between growth rates of the optimal gambler's wealth with
and without side information $Y$. Thus, Kelly's result gives an
interpretation that mutual information $I(X;Y)$ is the value of side
information $Y$ for the horse race $X$.

In order to tackle problems arising in information systems with
causally dependent components, Massey \cite{Massey90}
introduced the notion of directed information as
\begin{equation*}
I(X^n\to Y^n) \triangleq \sum_{i=1}^n I(X^i;Y_i|Y^{i-1}),
\end{equation*}
and showed that the maximum directed information upper bounds the
capacity of channels with feedback.  Subsequently, it was shown that
Massey's directed information and its variants indeed characterize the
capacity of feedback and two-way
channels~\cite{Kramer98,Tatikonda00,Kramer03,Permuter06_feedback_submit,Tatikonda06,Kim07_feedback,Permuter_Weissman_MAC07,ShraderPemuter07ISIT}
and the rate distortion function with feedforward
\cite{Pradhan07Venkataramanan}.

The main contribution of this paper is showing that directed
information $I(Y^n \to X^n)$ has a natural interpretation in gambling
as the difference in growth rates due to {\it causal} side
information. As a special case, if the horse race outcome and the
corresponding side information sequences are i.i.d., then the
(normalized) directed information becomes a single letter mutual
information $I(X;Y)$, and it coincides with Kelly's result.

The paper is organized as follows. We describe the notation of
directed information and causal conditioning in Section~\ref{sec:
directed information}. In Section~\ref{sec_problem_form}, we formulate
the horse-race gambling problem, in which side information is revealed
causally to the gambler. We present the main result in
Section~\ref{sec_main_res} and an analytically solved example in
Section~\ref{s_example}.  Finally, Section~\ref{s_conclusion}
concludes the paper and states two possible extensions of this work to
stock market and data compression with causal side information.

\section{Directed information and causal conditioning}
\label{sec: directed information}
Throughout this paper, we use the {\it causal conditioning} notation
$(\cdot||\cdot)$ developed by Kramer~\cite{Kramer98}. We denote as
$p(x^n||y^{n-d})$ the probability mass function (pmf) of $X^n =
(X_1,\ldots, X_n)$ \emph{causally conditioned} on $Y^{n-d}$, for some
integer $d\geq 0$, which is defined as
\begin{equation*} 
p(x^n||y^{n-d})\triangleq \prod_{i=1}^{n} p(x_i|x^{i-1},y^{i-d}).
\end{equation*}
(By convention, if $i-d \leq 0$ then $x^{i-d}$ is set to null.)  In
particular, we use extensively the cases $d=0,1$:
\begin{align*}
p(x^n||y^{n}) &\triangleq \prod_{i=1}^{n} p(x_i|x^{i-1},y^{i}),\\
p(x^n||y^{n-1}) &\triangleq \prod_{i=1}^{n} p(x_i|x^{i-1},y^{i-1}).
\end{align*}
Using the chain rule, we can easily verify that
\begin{equation*}
p(x^n,y^n)=p(x^n||y^{n})p(y^n||x^{n-1}).
\end{equation*}

The \emph{causally conditional entropy} $H(X^n||Y^n)$ is defined as
\begin{align*}
H(X^n||Y^n) &\triangleq \E[\log p(X^n||Y^n)] \notag\\
&=\sum_{i=1}^n H(X_i|X^{i-1},Y^i).
\end{align*}
Under this notation, directed information can be written as
\begin{align*}
I(Y^n \to X^n) &= \sum_{i=1}^n I(X_i;Y^i|X^{i-1}) \notag\\
&= H(X^n) - H(X^n||Y^n),
\end{align*}
which hints, in a rough analogy to mutual information, a possible
interpretation of directed information $I(Y^n \to X^n)$ as the amount
of information causally available side information $Y^n$ can provide
about $X^n$.

Note that the channel capacity results involve the term $I(X^n\to
Y^n)$, which measure the information in the forward link $X^n \to
Y^n$.  In contrast, in gambling the gain in growth rate is due to
the side information (backward link), and therefore the expression
$I(Y^n\to X^n)$ appears.

\section{Gambling in horse races with causal side information}
\label{sec_problem_form} Suppose that there are $m$ racing horses
in an infinite sequence of horse races and let $X_i \in \mathcal{X}
\triangleq [1,2,...,m],$ $i=1,2,\ldots,$ denote the horse that wins at
time $i$. Before betting in the $i$-th horse race, the gambler knows
some side information $Y_i \in \mathcal{Y}$. We assume that the
gambler invests all his capital in the horse race as a function of the
information that he knows at time $i$, i.e., the previous horse race
outcomes $X^{i-1}$ and side information $Y^i$ up to time $i$.  Let
$b(x_i|x^{i-1},y^{i})$ be the proportion of wealth that the gambler
bets on horse $x_i$ given $X^{i-1} = x^{i-1}$ and $Y^i = y^{i}$. The
betting scheme should satisfy $b(x_i|x^{i-1},y^{i})\geq 0$ (no short)
and $\sum_{x_i} b(x_i|x^{i-1},y^{i})=1$ for any history
$x^{i-1},y^i$. Let $o(x_i|x^{i-1})$ denote the odds of a horse $x_i$
given the previous outcomes $x^{i-1}$, which is the amount of capital
that the gambler gets for each unit capital invested in the horse. We
denote by $S(x^n||y^n)$ the gambler's wealth after $n$ races where the
race outcomes were $x^n$ and the side information that was causally
available was $y^n$. The {\it growth}, denoted by $W(X^n||Y^n)$, is
defined as the expected logarithm (base 2) of the gambler's wealth,
i.e.,
\begin{equation}\label{e_growth_rate_def}
W(X^n||Y^n)\triangleq \E[\log S(X^n||Y^n)].
\end{equation}
Finally the {\it growth rate}
$\frac{1}{n}W(X^n||Y^n)$ is defined as the normalized growth.

Here is a summary of the notation:
\begin{itemize}
\item $X_i$ is the outcome of the horse race at time $i$.
\item $Y_i$ is the the side information at time $i$.
\item ${o}(X_i|X^{i-1})$ is the payoffs at time $i$ for horse $X_i$ given that in the previous race the horses
$X^{i-1}$ won.
\item $b(X_i|Y^i,X^{i-1})$
the fractions of the gambler's wealth invested in horse $X_i$
at time $i$ given that the outcome of the previous races are
$X^{i-1}$ and the side information available at time $i$ is $Y^i$.
\item $S(X^n||Y^n)$ the gambler's wealth after $n$ races when the outcomes of the races are $X^n$ and the side information
$Y^{n}$ is causally available.
\item $\frac{1}{n}W(X^n||Y^n)$ is the growth rate.
\end{itemize}

Without loss of generality, we assume that the gambler's capital is 1
initially; therefore $S_0=1$.

\section{Main Results}
\label{sec_main_res} In  Subsection \ref{sub_all_money}, we assume
that the gambler invests all his money in the horse race while in
Subsection \ref{sub_part_of_the_mony}, we allow the gambler to
invest only part of the money. Using Kelly's result, it is shown in
Subsection \ref{sub_part_of_the_mony} that if the odds are fair with
respect to some distribution then the gambler should invest all his
money in the race.

\subsection{Investing all the money in the horse
race}\label{sub_all_money}
We assume that at any time $n$ the
gambler invests all his capital and therefore
\begin{equation*}
S(X^n||Y^n)\!=\!b(X_{n}|X^{n-1}\!\!,Y^{n})o(X_n|X^{n-1})S(X^{n-1}||Y^{n-1}).
\end{equation*}
This also implies that
\begin{equation*}
S(X^n||Y^n)=\prod_{i=1}^n b(X_{i}|X^{i-1},Y^{i})o(X_i|X^{i-1}).
\end{equation*}
The following proposition characterizes the optimal betting strategy
and the corresponding growth of wealth.

\medskip
\begin{theorem}\label{t_gamble_all_money}
For any finite horizon $n$, the maximum growth rate is achieved
when the gambler invests the money proportional to the causal
conditioning distribution, i.e.,
\begin{equation}\label{e_b=p}
b^*(x_i|x^{i-1},y^{i})=p(x_i|x^{i-1},y^{i}),
\quad \forall x^i, y^i, i\leq n,
\end{equation}
and the growth is
\begin{equation*}
W^*(X^n||Y^n)= \E[\log o(X^n)]-H(X^n||Y^n).
\end{equation*}
\end{theorem}

\medskip
Note that the sequence $\{p(x_i|x^{i-1},y^{i})\}_{i=1}^n$ uniquely
determines $p(x^n||y^{n})$. Also for all pairs $(x^n,y^{n})$ such that
$p(x^n||y^{n})>0$, the sequence $\{p(x_i|x^{i-1},y^{i})\}_{i=1}^n$ is
determined uniquely by $p(x^n||y^{n})$ simply by the identity
\begin{equation*}
p(x_i|x^{i-1},y^{i})=\frac{p(x^{i}||y^{i})}{p(x^{i-1}||y^{i-1})}.
\end{equation*}
A similar argument applies for $\{b^*(x_i|x^{i-1},y^{i})\}_{i=1}^n$
and $b^*(x^n||y^{n})$, and therefore (\ref{e_b=p}) is equivalent to
\begin{equation*}
b^*(x^n||y^n)=p(x^n||y^n),\ \ \forall x^n  \in \mathcal X^n, y^n\in
\mathcal Y^n.
\end{equation*}

\begin{IEEEproof}[Proof of Theorem~\ref{t_gamble_all_money}]
We have
\begin{align*}
W^*(X^n||Y^n)
&=\max_{b(x^{n}||y^n)} \E[\log
b(X^n||Y^n)o(X^n)]\nonumber \\
&= \max_{b(x^{n}||y^n)} \E[\log
b(X^n||Y^n)]+  \E[\log o(X^n)]\nonumber \\
&=-H(X^n||Y^n)+ \E[\log o(X^n)],
\end{align*}
where the last equality is achieved by choosing
$b(x^{n}||y^n)=p(x^{n}||y^n),$ and it is justified by the following
upper bound
\begin{align}\label{e_Elogb}
\E[&\log b(X^n||Y^n)]\nonumber \\
&=\sum_{x^n,y^n
}p(x^n,y^n)\left[ \log p(x^n||y^n)+\log\frac{b(x^n||y^n)}{p(x^n||y^n)}\right] \nonumber \\
&=-H(X^n||Y^n)+\sum_{x^n,y^n }p(x^n,y^n)\log\frac{b(x^n||y^n)}{p(x^n||y^n)}\nonumber \\
&\stackrel{(a)}{\leq} -H(X^n||Y^n)+\log\sum_{x^n,y^n
}p(x^n,y^n)\frac{b(x^n||y^n)}{p(x^n||y^n)}\nonumber \\
&\stackrel{(b)}{\leq} -H(X^n||Y^n)+\log\sum_{x^n,y^n }p(y^n||x^{n-1})b(x^n||y^n)\nonumber \\
&= -H(X^n||Y^n),
\end{align}
where (a) follows from Jensen's inequality and (b) from the fact that
$\sum_{x^n,y^n }p(y^n||x^{n-1})b(x^n||y^n)=1$. All summations in
\eqref{e_Elogb} are over the arguments $(x^n,y^n)$ for which
$p(x^n,y^n)>0$. This ensures that $p(x^n||y^n)>0$, and therefore, we
can multiply and divide by $p(x^n||y^n)$ in the first step
of \eqref{e_Elogb}.
\end{IEEEproof}

\medskip
In the case that the odds are fair and uniform, i.e.,
${o}(X_i|X^{i-1})=\frac{1}{|\mathcal X|}$, then
\begin{equation*}
\frac{1}{n}W^*(X^n||Y^n)= \log |\mathcal
X|-\frac{1}{n}H(X^n||Y^n).
\end{equation*}
Thus the sum of the growth rate $\frac{1}{n} W(X^n||Y^n)$ and the
entropy rate $\frac{1}{n} H(X^n||Y^n)$ of the horse race process
conditioned causally on the side information is constant, and one can
see a duality between $H(X^n||Y^n)$ and $W^*(X^n||Y^n)$;
cf.~\cite[th.~6.1.3]{CovThom06}

Let us denote by  $\Delta W$ the increase in the growth rate due
to causal side information, i.e.,
\begin{equation}
\Delta W=\frac{1}{n}W^*(X^n||Y^n)-\frac{1}{n} W^*(X^n).
\end{equation}
Thus $\Delta W$ characterizes the value of side information
$Y^n$. Theorem~\ref{t_gamble_all_money} leads to the following
proposition, which gives a new operational meaning of Massey's
directed information.
\begin{corollary}
The increase in growth rate due to causal side information $Y^n$ for
horse races $X^n$ is
\begin{equation}
\Delta W=\frac{1}{n}I(Y^n\to X^n).
\end{equation}
\end{corollary}
\begin{IEEEproof}
From Theorem \ref{t_gamble_all_money}, we have
\begin{align*}
W^*(X^n||Y^n)-W^*(X^n)&=-H(X^n||Y^n)+H(X^n) \nonumber \\
&= I(Y^n\to X^n). \tag*{\IEEEQED}
\end{align*}
\renewcommand{\IEEEQED}{}
\end{IEEEproof}
\renewcommand{\IEEEQED}{\IEEEQEDclosed}

\subsection{Investing only part of the
money}\label{sub_part_of_the_mony}
In this subsection we consider the case where the gambler does not
necessarily invest all his money in the gambling. Let
$b_0(y^i,x^{i-1})$ be the portion of money that the gambler does
not invest in gambling at time $i$ given that the previous races
results were $x^{i-1}$ and the side information is $y^i$. In this
setting, the wealth is given by
\begin{align*}
S&(X^n||Y^n) \nonumber\\
&=\prod_{i=1}^n \bigl(
b_0(X^{i-1},Y^{i})+(b(X_{i}|X^{i-1},Y^{i})o(X_i|X^{i-1})\bigr),\nonumber
\end{align*}
and the growth $W(X^n||Y^n)$ is defined as before in
(\ref{e_growth_rate_def}).

The term $W(X^n||Y^n)$ obeys a chain rule similar to the causal
conditioning entropy definition $H(X^n||Y^n)$, i.e.,
\begin{equation*}
W(X^n||Y^n)=\sum_{i=1}^n W(X_i|X^{i-1},Y^{i}),
\end{equation*}
where
\begin{align*} \lefteqn{ W(X_i|X^{i-1},Y^{i-1})}\nonumber \\
&\triangleq& E\left[\log
(b_0(X^{i-1},Y^i)+b(X_i|X^{i-1},Y^i)o(X_i|X^{i-1}))\right].\nonumber
\end{align*}
Note that for any given history $(x^{i-1},y^i) \in \mathcal{X}^{i-1}
\times \mathcal{Y}^i$, the betting scheme
$\{b_0(x^{i-1},y^i),b(x_i|x^{i-1},y^i) \}$ influences only
$W(X_i|X^{i-1},Y^{i})$, so that we have
\begin{align*}
&\max_{\{b_0(x^{i-1},y^i),b(x_i|x^{i-1},y^i)\}_{i=1}^n}W(X^n||Y^n)\nonumber \\
&=\sum_{i=1}^n
\max_{b_0(x^{i-1},y^i),b(x_i|x^{i-1},y^i)}
W(X_i|X^{i-1},Y^{i})\nonumber \\
&=\sum_{i=1}^n \sum_{x^{i-1},y^{i}}p(x^{i-1},y^i)
\!\!\!\!\maxx_{b_0(x^{i-1},y^i),b(x_i|x^{i-1},y^i)}\!\!\!\!
W(X_i|x^{i-1},y^{i}).
\end{align*}
The optimization problem in the last equation
is equivalent to the problem of finding the
optimal betting strategy in the memoryless case where the winning
horse distribution $p(x)$ is $p(x)=\Pr(X_i=x|x^{i-1},y^{i})$, the
odds $o(x)$ are $o(x)=o(X_i=x|x^{i-1})$, and the betting strategy
$(b_0,b(x))$ is $(b_0(x^i,y^{i-1}),b(X_i=x|x^{i-1},y^i))$,
respectively. Hence, the optimization, $\max
W(X_i|x^{i-1},y^{i})$, is equivalent to the following convex
problem:
\begin{align*}
\text{maximize } &\quad \sum_x p(x) \log(b_0+b(x)o(x))\nonumber \\
\text{subject to } &\quad b_0+\sum_x b(x)=1, \nonumber \\
&\quad b_0\geq 0, \ b(x)\geq0, \quad \forall x\in \mathcal X.
\end{align*}

The solution to this optimization problem was given by Kelly
\cite{Kelly56}. If the odds are \emph{super-fair}, namely, $\sum_x
\frac{1}{o(x)}\leq 1$, then the gambler will invest all his wealth in
the race rather than leave some as cash, since by betting
$b(x)=\frac{c}{o(x)}$, where $c=1/{\sum_x \frac{1}{o(x)}}$, the
gambler's money will be multiplied by $c\geq 1$, regardless of the
race outcome.  Therefore, for this case, the solution is given by
Theorem \ref{t_gamble_all_money}, where the gambler invests
proportional to the causal conditioning distribution $p(x^n||y^n)$.

If the odds are sub-fair, i.e., $\sum_x \frac{1}{o(x)}> 1$, then it is
optimal to bet only some of the money, namely $b_0>0$. The solution to
this problem is given in terms of an algorithm in
\cite[p. 925]{Kelly56}.

%

\section{An example}
\label{s_example} Here we consider betting in a horse race, where
the wining horse can be represented as a  Markov process, and
causal side information is available.
\begin{example}\label{ex1}
Consider the horse race process depicted in Figure~\ref{f_Markov}
where two horses are racing and the winning horse $X_i$ behaves as a
Markov process.  A horse that won will win again with probability
$1-p$ and lose with probability $p$. At time zero, we assume that both
horses have probability $\frac{1}{2}$ of wining. The side information
$Y_i$ at time $i$ is a noisy observation of the horse race outcome
$X_i$. It has probability $1-q$ of being equal to $X_i$, and
probability $q$ of being different from $X_i$.

For this example, the increase in growth rate due to side
information as $n$ goes to infinity is
\begin{equation*}
\Delta W =  h(p*q)-h(q),
\end{equation*}
where the function $h(\cdot)$ denotes the binary entropy, i.e.,
$h(x)=-x\log x-(1-x)\log (1-x),$ and $p*q$ denotes the parameter
of a Bernoulli distribution that results from convolving two
Bernoulli distributions with parameters $p$ and $q$, i.e.,
$p*q=(1-p)q+(1-q)p$.

The increase in the growth rate $\Delta W$ for this example can be
obtained using first principles as follows:
{\allowdisplaybreaks
\begin{align}\label{e_markov_qp}
&{\Delta W}\nonumber \\
&= \lim_{n\to \infty} \frac{1}{n} I(Y^n\to X^n) \nonumber \\
&= \lim_{n\to \infty} \frac{1}{n} \sum_{i=1}^n
H(Y^i|X^{i-1})-H(Y^i|X^{i})\nonumber \\
&= \lim_{n\to \infty} \frac{1}{n} \sum_{i=1}^n
\left[H(Y^i|X_{i-1})-H(Y_2^i|X_2^{i})-H(Y_1|X_1)\right] \nonumber \\
&\stackrel{\!(a)\!}{=} \!\lim_{n\to \infty} \frac{1}{n}
\sum_{i=1}^n
\left[H(Y^i|X_{i-1})-H(Y^{i-1}|X^{i-1})-H(Y_1|X_1)\right] \nonumber \\
&= \lim_{n\to \infty} \frac{1}{n} \sum_{i=1}^n
\left[ H(Y_i|Y^{i-1},X_{i-1})-H(Y_1|X_1)\right]\nonumber \\
&\stackrel{(b)}{=} H(Y_1|X_{0})-H(Y_1|X_1)
= h(p*q)-h(q),
\end{align}
where steps (a) and (b) are due to the stationarity of the process
$(X_i,Y_i)$.} Alternatively, the sequence of equalities up to step (b)
in (\ref{e_markov_qp}) can be derived directly using
\begin{align}\label{e_kim}
\frac{1}{n} I(Y^n\to X^n)
&\stackrel{(a)}{=} \frac{1}{n}\sum_{i=1}^n
I(Y_i;X_i^n|X^{i-1},Y^{i-1}) \nonumber \\
&\stackrel{(b)}{=} H(Y_1|X_{0})-H(Y_1|X_1),
\end{align}
where (a) is the identity given in \cite[eq.~(9)]{Kim07_feedback} and
(b) is due to the stationarity of the process.

If the side information is known with some lookahead $k\in
\{0,1,...\}$, that is, if the gambler knows $Y^{i+k}$ at time $i$,
then the increase in growth rate is given by
\begin{align}
\Delta W&= \lim_{n\to \infty} \frac{1}{n} I(Y^{n+k}\to X^n) \nonumber \\
&\stackrel{}{=} H(Y_{k+1}|Y^{k},X_{0})-H(Y_1|X_1),
\end{align}
where the last equality is due to the same arguments as in
(\ref{e_kim}).

\begin{figure}[t]{\footnotesize
\psfrag{Horse}[][][1]{Horse} \psfrag{Wins}[][][1]{wins}
\psfrag{1}[][][1]{$1$} \psfrag{2}[][][1]{$2$}
\psfrag{P}[][][1]{$p$} \psfrag{a}[][][1]{$1\!-\!p$}
\psfrag{Y1}[][][1]{$Y^n$} \psfrag{G1}[][][1]{}
\psfrag{G2}[][][0.9]{$X\!=\!1$} \psfrag{G3}[][][1]{$$}
\psfrag{T1}[][][1]{Horse 1 wins}
\psfrag{H2}[][][0.9]{$X\!=\!2$}
\psfrag{T2}[][][1]{Horse 2 wins}
\psfrag{c}[][][1]{1}
\psfrag{d}[][][1]{2}
\psfrag{X}[][][1]{$X$} \psfrag{Y}[][][1]{$Y$} \psfrag{T3}[][][1]{}
\psfrag{T4}[][][1]{} \psfrag{q}[][][1]{$q$}
\psfrag{q2}[][][1]{$1-q$}

\centerline{\includegraphics[width=6.6cm]{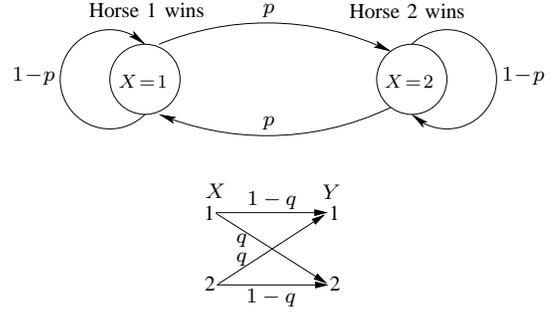}}
\caption{The setting of Example 1. The winning horse $X_i$ is
represented as a Markov process with two states. In state 1, horse
number 1 wins, and in state 2, horse number 2 wins. The side
information, $Y_i$, is noisy observation of the wining horse, $X_i$. }
\label{f_Markov} }
\end{figure}

\begin{figure*}[t]{
\psfrag{increase in growth rate}[][][1]{$\Delta W$}
\psfrag{q}[][][1]{$q$} \psfrag{n}[][][1]{$k$} \psfrag{a}[][][1]{$\
\ \ \rightarrow$} \psfrag{c}[][][0.8]{$H(X_1|X_0)\ \ \ \ \ $}
\psfrag{n}[][][1]{$k$} \psfrag{d}[][][0.8]{$\ \ \ \ \ \ \  \ \ \ \
\ \ \ \ \ \ \ H(Y_1|Y_0,Y_{-1},...)$} \psfrag{e}[][][0.8]{$\ \ \ \
\ \ \ \ \ \ \ \ -H(Y_1|X_1)$} \psfrag{b}[][][1]{$\leftarrow\ $}

\psfrag{n}[][][1]{$k$}
\centerline{\includegraphics[width=13.5cm]{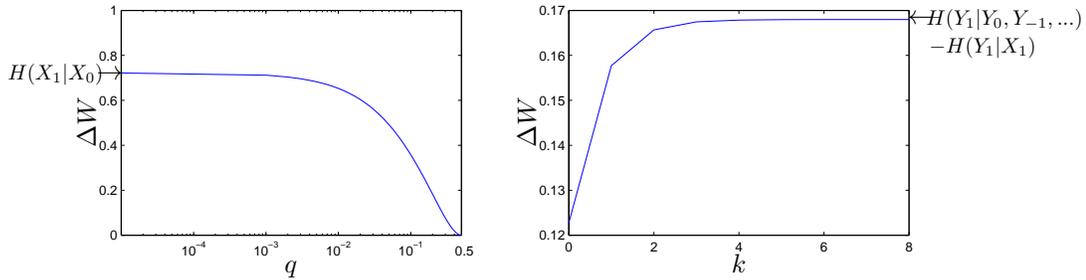}}
\caption{Increase in the growth rate, in Example \ref{ex1}, as a
function of the side information parameters $(q,k)$. The left plot of
the figure shows the increase of the growth rate $\Delta W$ as a
function of $q=\Pr(X_i\neq Y_i)$ and no lookahead. The right plot
shows the increase of the growth rate as function of lookahead $k$,
where $q=0.25$. The horse race outcome is assumed to be a first-order
binary symmetric Markov process with parameter $p=0.2$.}
\label{f_res}
}\end{figure*} 

Figure~\ref{f_res} shows the increase in growth rate $\Delta W$ due to
side information as a function of the side information parameters
$(q,k)$. The left plot shows $\Delta W$ as a function of $q$,
where $p=0.2$ and no lookahead, $k=0$. The right plot shows
$\Delta W$ as a function of $k$, where $p=0.2$ and $q=0.25$. If the
entire side information sequence ${Y_1,Y_2,...}$ is known to the
gambler ahead of time, then we should have mutual information rather
then directed information, i.e.,
\begin{align}
\Delta W&= \lim_{n\to \infty} \frac{1}{n} I(Y^{n};X^n)\nonumber
\\
&=\lim_{n\to \infty} \frac{H(Y^n)}{n} -H(Y_1|X_1),
\end{align}
and this coincides with the fact that for a stationary hidden
Markov process $\{Y_1,Y_2,...\}$ the sequence
$H(Y_{k+1}|Y^{k-1},X_{0})$ converges to the entropy rate of the
process.
\end{example}

\section{Conclusion and further extensions}
\label{s_conclusion} We have shown that directed information arises
naturally in gambling as the gain in the maximum achievable capital
growth due to the availability of causal side information. We now
outline two extensions: stock market portfolio strategies and data
compression in the presence of causal side information. Details are
given in \cite{KPW08}.

\subsection{Stock market}
Using notation similar to that in \cite[ch. 16]{CovThom06}, a stock
market at time $i$ is represented as a vector of stocks ${\bf
X}_i=(X_{i1},X_{i2},...,X_{im})$, where $m$ is the number of
stocks, and the {\it price relative $X_{ik}$} is the ratio of the
price of stock-$k$ at the end of day $i$ to the price of stock-$k$ at
the beginning of day $i$. We assume that at time $i$ there is side
information $Y^i$ that is known to the investor. 
A {\it portfolio} 
 is an allocation of wealth across the stocks. 
A nonparticipating or causal portfolio strategy with causal side
information at time $i$ is denoted as ${\bf b}({\bf
x}^{i-1},y^i)$, and it 
satisfies $\sum_{k=1}^m b_{k}({\bf x}^{i-1},y^i) =1$, and $b_{k}({\bf
X}^{i-1},Y^i)\geq 0$ for all possible ${\bf x}^{i-1},y^i$. We define
$S({\bf x}^n||y^n)$ as the wealth at the end of day $n$ for a stock
sequence ${\bf x}^n$ and causal side information $y^n$. We can write
\begin{equation*}
S({\bf x}^n||y^n) = \left ({\bf b}^t({\bf x}^{n-1},y^n) {\bf
x}_n\right ) S({\bf x}^{n-1}||y^{n-1})
\end{equation*}
where $(\cdot)^t$ denotes the transpose of a vector.
The goal is to maximize the growth $ W({\bf X}^n||Y^n)=\E[\log S({\bf
X}^n||Y^n) ].$ We also define
$
W({\bf X}_n|{\bf X}^{n-1},Y^n)=\E[\log ({\bf b}^t({\bf X}^{n-1},Y^n)
{\bf X}_n)].
$
From this definition, we can write the chain rule
\begin{equation*}
W({\bf X}^n||Y^n)=\sum_{i=1}^n W({\bf X}_i|{\bf X}^{i-1},Y^i).
\end{equation*}

The gambling in horse races with $m$ horses studied in the previous
section is a special case of investing the stock market with $m+1$
stocks. The first $m$ stocks correspond to the $m$ horses and at the
end of the day one of the stocks, say $k\in \{1,...,m\}$, gets the
value $o(k)$ with probability $p(k)$ and all other stocks become
zero. The $m+1$-st stock is always one, and it allows the gambler to
invest only part of the wealth in the horse race.

The developments in the previous section can be expanded to
characterize the increase in growth rate due to side information,
where again directed information emerges as the key quantity,
upper-bounding the value of causal side information;
cf.\@~\cite{BarCov1988}. Details will be given in \cite{KPW08}.

\subsection{Instantaneous  compression with causal side information}

Let $X_1, X_2, \ldots$ be a source and $Y_1, Y_2, \ldots$ its side
information sequence. The source is to be losslessly encoded
instantaneously, with causal available side information. More
precisely, an instantaneous lossless source encoder with causal side
information
consists of a sequence of mappings $\{ M_i \}_{i \geq 1}$ such that
each $M_i: \mathcal{X}^i \times \mathcal{Y}^i \to \{0,1\}^*$ has the
property that for every $x^{i-1}$ and $y^i$ $M_i (x^{i-1}\; \cdot \,,
y^i)$ is an instantaneous (prefix) code for $X_i$.

An instantaneous lossless source encoder with causal side information
operates sequentially, emitting the concatenated bit stream $M_1 (X_1,
Y_1) M_2(X^2, Y^2) \cdots$. The defining property that $M_i (x^{i-1}
\cdot , y^i)$ is an instantaneous code for every $x^{i-1}$ and $y^i$
is a necessary and sufficient condition for the existence of a decoder
that can losslessly recover $x^i$ based on $y^i$ and the bit stream
$M_1 (x_1, y_1) M_2(x^2, y^2) \cdots$ just as soon as it sees $M_1
(x_1, y_1) M_2(x^2, y^2) \cdots M_i(x^i, y^i)$, for all sequence pairs
$(x_1,y_1), ( x_2, y_2) \ldots$ and all $i \geq 1$.  Using natural
extensions of standard arguments we show in \cite{KPW08} that $I(Y^n
\rightarrow X^n)$ is essentially (up to terms that are sublinear in
$n$) the rate savings in optimal sequential lossless compression of
$X^n$ due to the causal availability of the side information.

\bibliographystyle{IEEEtran}
\bibliography{isit3}

\end{document}